# Novel Efficient Scalable QCA XOR and Full Adder Designs

Behrouz Safaiezadeh[1], Majid Haghparast[2, *] and Lauri Kettunen[2]

[1] Department of Computer, Andimeshk Branch, Islamic Azad University, Andimeshk, Iran

[2] Faculty of Information Technology, University of Jyväskylä, Jyväskylä, Finland

* Corresponding author: Majid Haghparast (e-mail: majid.m.haghparast@jyu.fi).

**Abstract:** Circuit design based on Quantum-dots Cellular Automata technology offers power-efficiency and nano-size circuits. It is an attractive alternative to CMOS technology. The XOR gate is a widely used building element in arithmetic circuits. An efficient XOR gate in QCA computational circuits can significantly improve efficiency. This paper proposes two different approaches for designing 3-input QCA XOR gates with 10 and 8 cells. They require two clock phases to create output. They have efficient and scalable structures. To demonstrate the functionality of these structures, we design QCA full adders using the suggested gates and compare the results with existing designs. The proposed QCA full adder has only 12 cells and is the best compared to all the existing counterparts. We simulated and verified the proposed structures. We proved the functionality of the proposed QCA full adder and the suggested QCA XOR structures. Additionally, QCAPro is used to estimate the energy dissipation of the proposed XOR and Full-adder. The results demonstrated that the proposed designs have the desired performance based on the number of cells, occupied area, and latency.

## 1. Introduction

QCA technology has many unique characteristics for designing logic circuits, including high speed, tiny size, and low power consumption. Many factors, like physical scalability, short channel effects, heating, and cooling issues, indicate that CMOS technology is reaching its limits [1]. Hence, QCA technology is a prominent alternative for the design of logic circuits. It uses square cells with four quantum dots arranged in each corner of the square cells. In such a quantum cell, the two electrons place separate positions based on the Columbia repulsion forces to minimize energy.

This paper presents two novel 3-input XOR gates in QCA. The proposed XOR gates consist of only 8 and 10 cells. They require two clock phases. In digital circuits, addition operations are necessary. In the literature, various adders for arithmetic operations can be found. The ongoing scientific research explores the design of adders with QCA technology to increase its performance by reducing circuit cells, occupied area, time delay, and overall circuit design costs.

To design an efficient QCA-based adder, we propose a new layout of full adders based on our suggested efficient XOR gates. We simulate the suggested structures with QCADesigner [2] and provide results that verify the function of the suggested QCA XOR and the QCA full adder. Also, we provide some output results of simulations to confirm the function of the suggested QCA XOR and the QCA full adder. Based on the results obtained, the proposed adders are efficient in almost all metrics, such as the area, cell count, and latency. Hence, the proposed structures are found to be prominent. In this paper, our primary objectives are:

- To present two novel XOR gates with 3-input in QCA and verify their function with simulations.
- To present two efficient full adders based on the proposed XOR gates in QCA and verify their function with simulations.
- To proposed XORs are scalable.
- To calculate the energy consumption of the proposed XOR and full-adder
- To compare the proposed XOR gates and full adders with designs suggested in the literature.

The structure of the rest of the paper is as follows: Section 2 reviews the previous designs of the XOR gates and full adders in QCA. Section 3 explains the suggested designs for the three input XOR gate and full adder. In Section 4, we provide simulation results and a comparison with existing counterparts. The final section presents a summary of the paper's achievements.

## 2. Basic concepts of QCA

A QCA cell is the basis of QCA technology. A quantum cell consists of four holes that exist in the corners. The distance between the cells is 2nm. In addition, each cell contains two additional electrons that can freely move between holes. The placement of two electrons in four quantum dots can take on six different states in general, but all these states are not stable. There is a Coulombic repulsion force between electrons, which causes them to always be in a state with the largest distance between them. Therefore, stable states occur when the electrons place diagonally. As seen in Fig. 1, these two states indicate the two polarities of +1 and -1, which have logic "1" and "0", respectively [3].

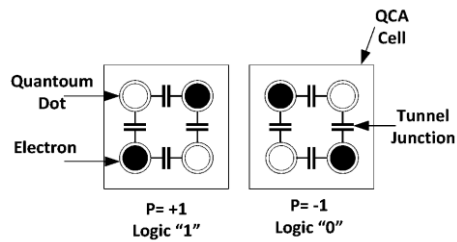

Fig.1. Two stable states of QCA cells with logic "1" and "0" [1]

A QCA cell array can transmit the information as a wire. Fig 2 depicts two types of QCA wires. These wires consist of normal cells (Fig. 2 a.) and rotate cells (Fig. 2 b).

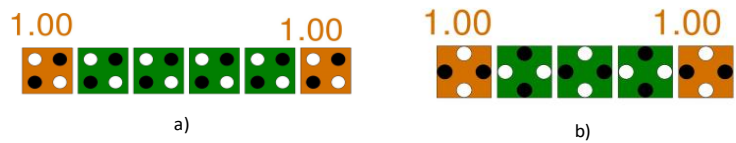

Fig. 2. QCA wire with a) Normal cells b) Rotate cells [2]

The majority and inverter gates are basic logic circuits in QCA. Figs. 3a and 3b show examples of the inverter and 3-Inputs Majority Voter (MV3) with QCA cells. The majority gate constructs logical AND, and OR gates if one of its inputs is zero and one (Figs. 3c and 3d). QCA technology can produce all combinational and sequential circuits using inverter and majority gates [5].

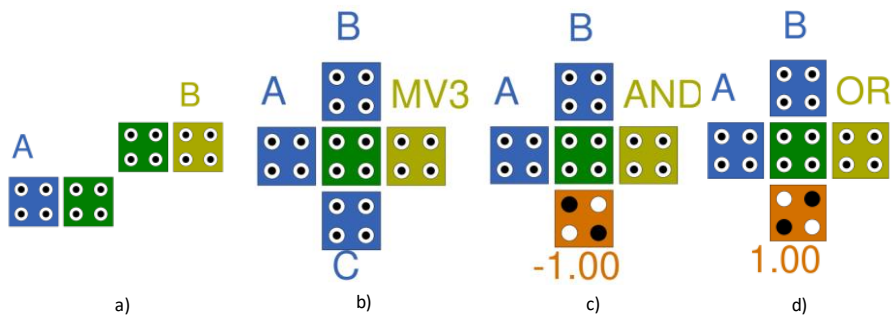

Fig. 3 Designing logical gates in QCA:    a) NOT    b) MV3    c) AND    d) OR [3]

## 3. State-of-the-art designs

In the literature, many structures for the full adder and XOR gate in QCA technology are proposed, and there is a lot of ongoing work around them. The full adder is a critical feature of the digital arithmetic unit, and its performance has an impact on the global system. Typically, the goal has been to develop optimal adders in terms of cell count, occupied area, and latency. We investigate previous full adder designs and XOR gates. Majeed et al. [4] proposed two structures of the

QCA XNOR. Then, they proposed a novel QCA 3-input XOR, which presented a single-bit comparator. Their QCA XOR included 11 cells, 0.01 $\mu m^2$ area, and two clock phases to produce output. Seyedi et al. [5] offered a full adder in QCA-based arithmetic units. Their proposed structure consists of two majority and two inverter gates. Majeed et al. [6] proposed a one-bit full adder that included only 15 cells, 0.007 $\mu m^2$ the area, and two clock phases to produce output. Safoev et. al. [7] proposed an XOR with three inputs that consist of 17 cells and two clock phases to produce output. Also, they suggested a full adder that had 35 cells. Salimzadeh et al. [8] proposed a new and efficient design for the XNOR in QCA technology to realize the implementation of a novel full-adder structure. Figs .4 and 5 show the QCA layout of the best previous works.

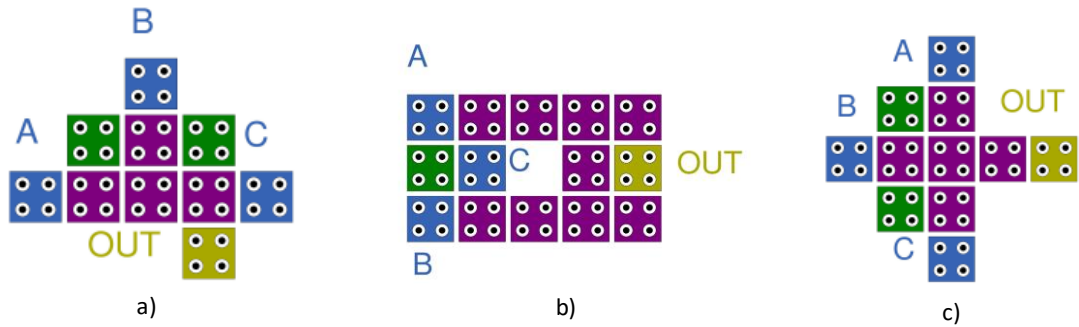

Fig. 4. QCA layout of different 3-input XOR: a) in [2] b) in [6] c) in [3]

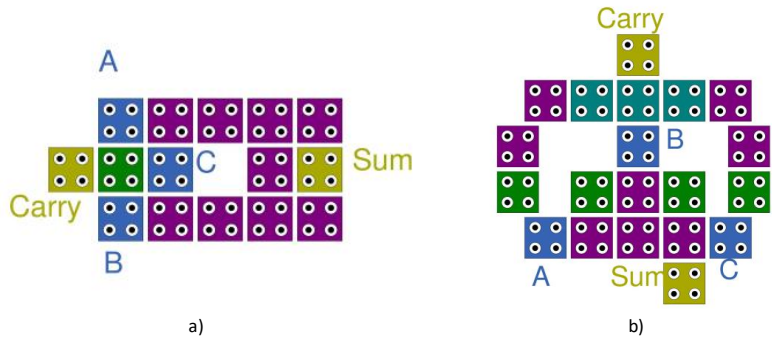

Fig. 5. QCA layout of different 3-input Full Adder: a) in [6] b) in [2]

4. Proposed structures

In this section, we present two new efficient and scalable QCA structures of the 3-input XOR gates. Then, we design new structures of full adder using the proposed XORs. All the proposed structures are single layer.

4.1 The Proposed Three-inputs QCA XOR Gates

The XOR gate is a base unit for designing arithmetic circuits. Although there have been numerous XOR layouts described earlier in QCA technology, our objective is to reduce the number of cells, area, and latency. We describe here the proposed QCA structures for the 3-input XOR gate. Fig. 6 illustrates the proposed structures of 3-inputs XOR. Both proposed structures are scalable. Scalability is a very important feature. The first design has only ten QCA cells, a latency of two clock phases, and only 0.005 μm² of area (Fig 6.a). The second design has only eight QCA cells, a latency of two clock phases, and only 0.0066 μm² of area (Fig 6.b). The second design has the lowest number of cells compared with the earlier proposed structures.

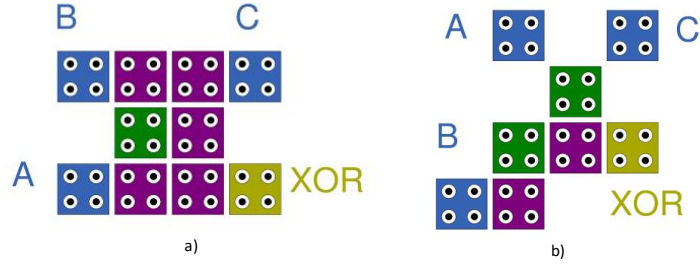

Fig. 6. Proposed structures of 3-input QCA XOR: a) First design b) Second design

Next, we simulate the second proposed gate to show its physical accuracy. To verify the gates, the kink energies from the input cells to output cells is computed. As the full simulations with all possible values are rather heavy, we calculate the value of the kink energy for A= 1, B= 0, and C=0. Fig. 7 displays the cellular network affecting the output cell with "+1" and "-1" polarities. By Eq. (1) [9], we compute the kink energy between two electron charges.

$$E_{ij} = \frac{k q_i q_j}{r_{ij}} = \frac{A}{r_{ij}}, \quad (1)$$

where $E_{ij}$ denotes kink energy, K is a constant ($k = 9 * 10^9$), $q_i$ and $q_j$ are electrical charges ($q = 1.6 * 10^{-19}$), and $r_{ij}$ denotes the distance between two electrical charges of i and j. Eq. (2) indicates the total electrostatic energy delivered to the $q_i$ electron.

$$U = \sum_{j=1}^{N} E_{ij} \quad (2)$$

The value of A is: $A = K * q_1 * q_2 = 9 * 10^9 * 1.6 * 10^{-19} * 1.6 * 10^{-19} = 23.04 * 10^{-29}$.
Therefore,

$$E_{ij} = \frac{A}{r_{ij}} = \frac{23.04 * 10^{-29}}{r_{ij}} \quad (3)$$

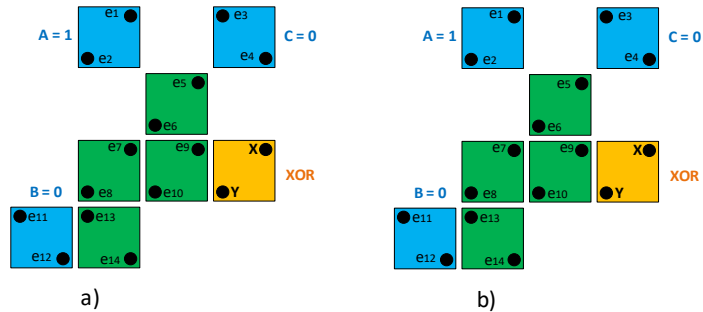

Fig. 7. The cellular network a) Polarity= +1 b) Polarity = -1

We calculate the kink energy between electrons X and Y of the output cell with other electrons, as shown in Tables 1 and 2. The calculated values in Table 1 are for the case where we assumed that the output of the structure is equal to one. In the left column in Table 1, we calculated the amount of the kink energy affected by all electrons on the electron at position X. In the right column in Table 1, we calculated the amount of the kink energy affected by all electrons on the electron at position Y.

Table 1. Calculation of kink energy between output cell electrons and all electrons in Fig. 7 a.

| For electron X | For electron Y |
|---|---|
| $U_{1X} = \dfrac{A}{r_{1x}} = \dfrac{23.04 * 10^{-29}}{56.57 * 10^{-9}} = 0.40 * 10^{-20}$ | $U_{1Y} = \dfrac{A}{r_{1Y}} = \dfrac{23.04 * 10^{-29}}{45.65 * 10^{-9}} = 0.50 * 10^{-20}$ |
| $U_{2X} = \dfrac{A}{r_{2x}} = \dfrac{23.04 * 10^{-29}}{62.03 * 10^{-9}} = 0.37 * 10^{-20}$ | $U_{2Y} = \dfrac{A}{r_{2Y}} = \dfrac{23.04 * 10^{-29}}{55.17 * 10^{-9}} = 0.42 * 10^{-20}$ |
| $U_{3X} = \dfrac{A}{r_{3x}} = \dfrac{23.04 * 10^{-29}}{43.86 * 10^{-9}} = 0.52 * 10^{-20}$ | $U_{3Y} = \dfrac{A}{r_{3Y}} = \dfrac{23.04 * 10^{-29}}{58 * 10^{-9}} = 0.40 * 10^{-20}$ |
| $U_{4X} = \dfrac{A}{r_{4x}} = \dfrac{23.04 * 10^{-29}}{22 * 10^{-9}} = 1.15 * 10^{-20}$ | $U_{4Y} = \dfrac{A}{r_{4Y}} = \dfrac{23.04 * 10^{-29}}{43.86 * 10^{-9}} = 0.53 * 10^{-20}$ |
| $U_{5X} = \dfrac{A}{r_{5x}} = \dfrac{23.04 * 10^{-29}}{28.28 * 10^{-9}} = 0.81 * 10^{-20}$ | $U_{5Y} = \dfrac{A}{r_{5Y}} = \dfrac{23.04 * 10^{-29}}{38.02 * 10^{-9}} = 0.61 * 10^{-20}$ |
| $U_{6X} = \dfrac{A}{r_{6x}} = \dfrac{23.04 * 10^{-29}}{38.05 * 10^{-9}} = 0.61 * 10^{-20}$ | $U_{6Y} = \dfrac{A}{r_{6Y}} = \dfrac{23.04 * 10^{-29}}{28.28 * 10^{-9}} = 0.81 * 10^{-20}$ |
| $U_{7X} = \dfrac{A}{r_{7x}} = \dfrac{23.04 * 10^{-29}}{40 * 10^{-9}} = 0.58 * 10^{-20}$ | $U_{7Y} = \dfrac{A}{r_{7Y}} = \dfrac{23.04 * 10^{-29}}{28.42 * 10^{-9}} = 0.81 * 10^{-20}$ |
| $U_{8X} = \dfrac{A}{r_{8x}} = \dfrac{23.04 * 10^{-29}}{60.27 * 10^{-9}} = 0.38 * 10^{-20}$ | $U_{8Y} = \dfrac{A}{r_{8Y}} = \dfrac{23.04 * 10^{-29}}{40 * 10^{-9}} = 0.58 * 10^{-20}$ |
| $U_{9X} = \dfrac{A}{r_{9x}} = \dfrac{23.04 * 10^{-29}}{20 * 10^{-9}} = 1.15 * 10^{-20}$ | $U_{9Y} = \dfrac{A}{r_{9Y}} = \dfrac{23.04 * 10^{-29}}{18.11 * 10^{-9}} = 1.27 * 10^{-20}$ |
| $U_{10X} = \dfrac{A}{r_{10x}} = \dfrac{23.04 * 10^{-29}}{42.047 * 10^{-9}} = 0.55 * 10^{-20}$ | $U_{10Y} = \dfrac{A}{r_{10Y}} = \dfrac{23.04 * 10^{-29}}{20 * 10^{-9}} = 1.15 * 10^{-20}$ |
| $U_{11X} = \dfrac{A}{r_{11x}} = \dfrac{23.04 * 10^{-29}}{80.52 * 10^{-9}} = 0.29 * 10^{-20}$ | $U_{11Y} = \dfrac{A}{r_{11Y}} = \dfrac{23.04 * 10^{-29}}{60.03 * 10^{-9}} = 0.38 * 10^{-20}$ |
| $U_{12X} = \dfrac{A}{r_{12x}} = \dfrac{23.04 * 10^{-29}}{71.02 * 10^{-9}} = 0.32 * 10^{-20}$ | $U_{12Y} = \dfrac{A}{r_{12Y}} = \dfrac{23.04 * 10^{-29}}{46.51 * 10^{-9}} = 0.49 * 10^{-20}$ |
| $U_{13X} = \dfrac{A}{r_{13x}} = \dfrac{23.04 * 10^{-29}}{61.35 * 10^{-9}} = 0.38 * 10^{-20}$ | $U_{13Y} = \dfrac{A}{r_{13Y}} = \dfrac{23.04 * 10^{-29}}{40.04 * 10^{-9}} = 0.58 * 10^{-20}$ |
| $U_{14X} = \dfrac{A}{r_{14x}} = \dfrac{23.04 * 10^{-29}}{43.86 * 10^{-9}} = 0.53 * 10^{-20}$ | $U_{14Y} = \dfrac{A}{r_{14Y}} = \dfrac{23.04 * 10^{-29}}{29.73 * 10^{-9}} = 0.84 * 10^{-20}$ |
| $U_X = \sum_{i=1}^{14} U_{iX} = 7.936 * 10^{-20}$ | $U_Y = \sum_{i=1}^{14} U_{iY} = 9.367 * 10^{-20}$ |
| $U = U_X + U_Y = (7.936 + 9.367) * 10^{-20} = 17.3 * 10^{-20}$ ||

The calculated values in Table 2 are for the case where we assumed that the output of the structure is equal to zero. Fig. 7.b shows the cellular network affecting the output cell with "-1" polarity. We calculated the amount of kink energy affected by all electrons on the electron at position X, and we reported it in the left column of Table 2. In the right column in Table 2, we calculated the amount of kink energy affected by all electrons on the electron at position Y. The simulation results for kink energy calculation are in the Appendix section. We provided detailed information about computing $U_X$ and $U_Y$ in the Appendix section in Table I.

Table 2. Calculation of kink energy between output cell electrons and all electrons in Fig. 7 b.

| For electron X | For electron Y |
|---|---|
| $U_X = \sum_{i=1}^{14} U_{iX} = 20.55 * 10^{-20}$ | $U_Y = \sum_{i=1}^{14} U_{iY} = 6.7 * 10^{-20}$ |
| $U = U_X + U_Y = (20.55 + 6.7) * 10^{-20} = 27.25 * 10^{-20}$ ||

We conclude from the results that the output cell electrons are located at position Fig. 7.a. Because it has less kink energy than Fig. 7.b. This means that the binary value "1" is the output cell polarity, which is correct.

**4.2 The Proposed QCA Full Adders**

Full adder has great importance as a building block of digital computation circuits. In related works, we reviewed previous designs. The most important issues with the existing QCA full adders are their cell count and occupied area.

This is due to using inefficient QCA XOR structures. Hence, we first suggested two structures for XORs in QCA technology that do not share the same issue. We will use our proposed QCA XORs in designing our suggested QCA full adders. The First full adder is designed based on the first of the proposed XOR. It contains only normal cells. Its structure includes 14 QCA cells, 0.01 area, and two clock phases to produce the output circuit. The second full adder is proposed based on the second proposed QCA XOR structure. This design includes 12 QCA cells, 0.01 area, and two clock phases to produce the output circuit. As a result, the suggested structures, with their particular properties, have the potential to alter computation at the nanoscale. Fig. 8 depicts the QCA structure of the proposed full adders.

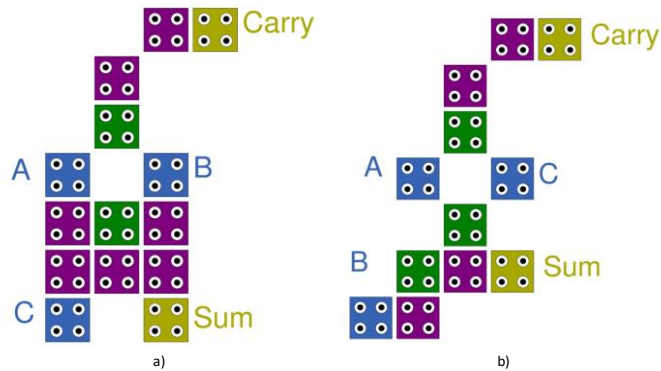

Fig. 8. The QCA structure of the proposed full adders a) First design b) Second design

The proposed full adder (Second design) is also simulated to determine its physical accuracy (See Fig. 8 b). For this, we need to compute the kink energies from the input cells to the output cells. We calculate the value of the kink energy for A = 1, B= 0, and C=1. Fig. 9 displays the cellular network affecting the output cells, which are called Carry and Sum with "+1" and "-1" polarities.

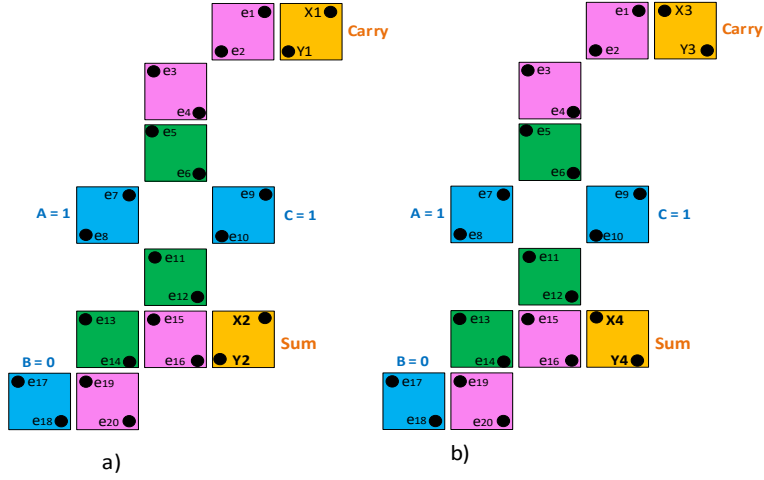

Fig. 9. The cellular network of the proposed full adder: a) Carry and Sum are one. b) Carry and Sum are Zero

We calculate the kink energy between electrons ($X_1$ and $Y_1$), ($X_2$ and $Y_2$), ($X_3$ and $Y_3$) and ($X_4$ and $Y_4$) of the output cells with other electrons, as shown in Tables 3, 4, 5, and 6, respectively. The simulation results for kink energy calculations are provided in the Appendix section in Tables II, III, IV, and V. The calculated values in Table 3 are for the case where we assumed that the output of the structure or Carry is one. In Table 4, the calculated values are for the case where we assumed that the output of the structure or Sum is one. The calculated values in Table 5 are for the case where we assumed that the output of the structure or Carry is zero. In Table 6, the calculated values are for the case where we assumed that the output of the structure or Sum is zero.

Table 3. Calculation of Kink energy between all electrons and $X_1$ and $Y_1$ electrons of the output cell of Fig. 6

| For electron $X_1$ | For electron $Y_1$ |
|---|---|
| $U_{X_1} = \sum_{i=1}^{20} U_{iX_1} = 6.065 * 10^{-20}$ | $U_{Y_1} = \sum_{i=1}^{20} U_{iY_1} = 8.57 * 10^{-20}$ |
| $U_{X_1Y_1} = U_{X_1} + U_{Y_1} = (6.065 + 8.57) * 10^{-20} = 14.635 * 10^{-20}$ ||

Table 4. Calculation of the Kink energy between all electrons with $X_2$ and $Y_2$ electrons of the output cell of Fig. 6

| For electron $X_2$ | For electron $Y_2$ |
|---|---|
| $U_{X_2} = \sum_{i=1}^{20} U_{iX_2} = 9.544 * 10^{-20}$ | $U_{Y_2} = \sum_{i=1}^{20} U_{iY_2} = 21.277 * 10^{-20}$ |
| $U_{X_2Y_2} = U_{X_2} + U_{Y_2} = (9.544 + 21.277) * 10^{-20} = 30.821 * 10^{-20}$ ||

Table 5. Calculation of the Kink energy between all electrons with $X_3$ and $Y_3$ electrons of the output cell of Fig. 6

| For electron $X_3$ | For electron $Y_3$ |
|---|---|
| $U_{X_3} = \sum_{i=1}^{20} U_{iX_3} = 17.324 * 10^{-20}$ | $U_{Y_3} = \sum_{i=1}^{20} U_{iY_3} = 6.524 * 10^{-20}$ |
| $U_{X_3Y_3} = U_{X_3} + U_{Y_3} = (17.324 + 6.524) * 10^{-20} = 23.848 * 10^{-20}$ ||

Table 6. Calculation of the Kink energy between all electrons with $X_4$ and $Y_4$ electrons of the output cell of Fig. 6

| For electron $X_4$ | For electron $Y_4$ |
|---|---|
| $U_{X_4} = \sum_{i=1}^{20} U_{iX_4} = 19.24 * 10^{-20}$ | $U_{Y_4} = \sum_{i=1}^{20} U_{iY_4} = 8.59 * 10^{-20}$ |
| $U_{X_4Y_4} = U_{X_4} + U_{Y_4} = (19.24 + 8.59) * 10^{-20} = 27.83 * 10^{-20}$ ||

A stable state has less energy. Therefore, by comparing values of the kink energy for output Carry, it concludes that the value of one is correct for it. Also, by comparing values of the kink energy for output Sum, it concludes that its value is zero.

## 5. Simulation Results, Discussion, and Power dissipation

We display simulation results of the proposed designs, compare them to previous works, and analyze the power dissipation of the proposed XOR and Full-Adder in this section.

### 5.1 Simulation Results and Discussion

The simulation results for the QCA XOR gates for all combinations of inputs are depicted in Fig. 10. Simulation results show that the proposed structures work well and perform effectively. Three input cells are A, B, and C. Output is cell XOR. When A=1, B=0, and C=1, the correct output is XOR=0.

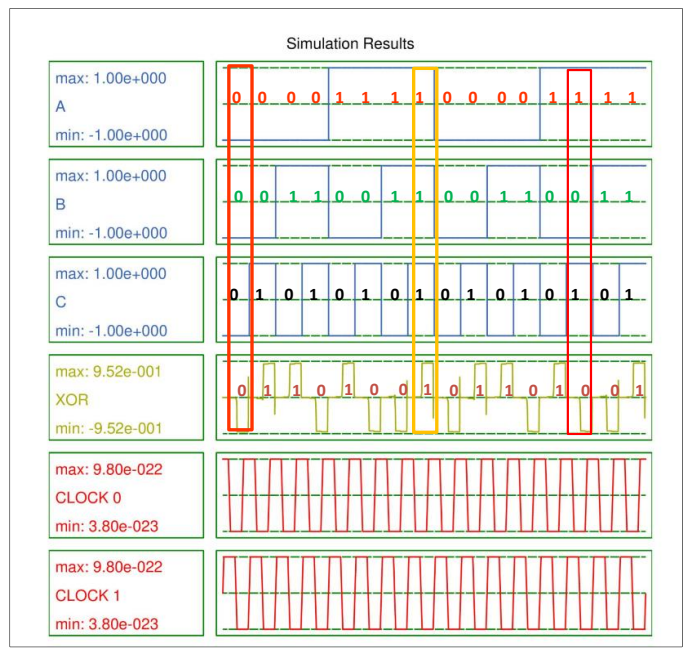

Fig.10. Simulation result of the proposed QCA XOR gates

Fig. 11 depicts the simulation results of the proposed full adders for all combinations of inputs. Simulation results show that the proposed structures work well and perform effectively. Three input cells are A, B, and C. Two output cells are Carry and Sum. When A=0, B=1, and C=1, the correct outputs are Carry=1 and Sum=0.

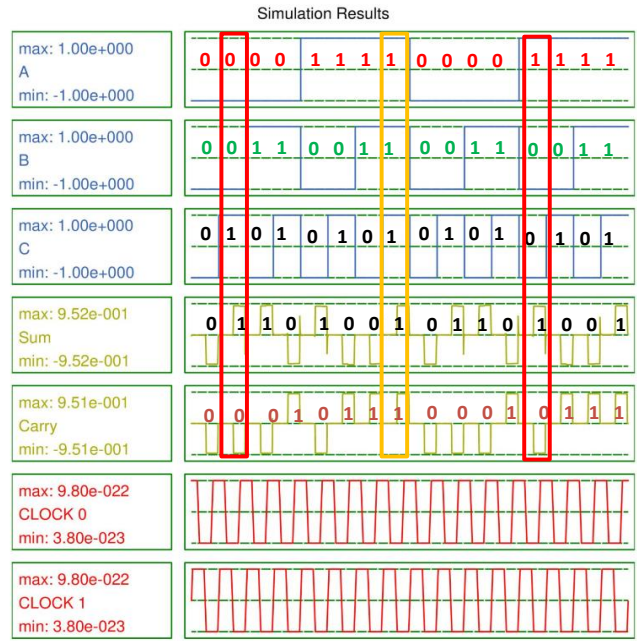

Fig. 11 Simulation result of the suggested one-bit full adder

To evaluate the proposed designs, we consider parameters of cell counts, occupied area, latency, cost, and the number of layers. The cost of the circuit is calculated based on Eq. (4) [10]. Table 7 compares the proposed QCA XOR gates and previous designs. It indicates that the proposed XOR outperforms the previous designs in terms of area, cell, and cost. The second design has the lowest number of cells compared to all the previous structures.

$$Cost = Cell\ count * Area * Latency \quad (4)$$

Table 7. Comparison of the 3-input QCA XOR gates

| Ref. | Cell count | Area ($\mu m^2$) | Latency (Number of Clocks) | Cost |
|---|---|---|---|---|
| Design in [11] | 93 | 0.07 | 1.25 | 8.137 |
| Design in [12] | 32 | 0.02 | 1 | 0.64 |
| Design in [13] | 30 | 0.0233 | 0.75 | 0.524 |
| Design in [14] | 28 | 0.02 | 0.75 | 0.42 |
| Design in [7] | 17 | 0.015 | 0.5 | 0.127 |
| Design in [6] | 14 | 0.006 | 0.5 | 0.042 |
| Design in [15] | 12 | 0.0116 | 0.5 | 0.0696 |
| Design in [16] | 11 | 0.01 | 0.5 | 0.055 |
| Design in [4] | 11 | 0.01 | 0.5 | 0.055 |
| Design in [2] | 10 | 0.01 | 0.5 | 0.05 |
| Our proposed (First approach) | 10 | 0.005 | 0.5 | 0.025 |
| Our proposed (Second approach) | 8 | 0.006 | 0.5 | 0.024 |

Table 8 compares the proposed full adders and previous designs. It indicates that the proposed full adder outperforms the previous designs in terms of the cell and cost. The second design has the lowest number of cells compared to all the previous structures. This design has approximately 21.5% improvement in cell numbers compared to the best existing QCA full adder. Also, the design surpasses all the previous robust QCA-based full adders in terms of area, delay, and complexity. The best previous full adder [6] has 15 QCA cells and an area of $2\mu m^2$. It has a more efficient area compared to the proposed designs. But it has some issues: The design in [6] is not scalable, and the structure surrounds one of its inputs. For using this structure in a modular way, it is necessary to design the circuit in a multilayered manner or to use rotating cells. The proposed design, consisting only of normal cells and of a single layer and solves such issues.

Table 8. Comparison of different QCA full adders

| Ref. | Cell count | Area ($\mu m^2$.) | Latency (Number of Clocks) | Cost |
|---|---|---|---|---|
| Design in [5] | 28 | 0.01 | 0.75 | 0.21 |
| Design in [17] | 22 | 0.01 | 0.75 | 0.165 |
| Design in [18] | 30 | 0.004 | 1 | 0.12 |
| Design in [3] | 37 | 0.04 | 0.75 | 1.11 |
| Design in [19] | 44 | 0.06 | 1.25 | 3.3 |
| Design in [20] | 36 | 0.04 | 0.75 | 1.08 |
| Design in [2] | 20 | 0.016 | 0.75 | 0.24 |
| Design in [21] | 37 | 0.025 | 0.75 | 0.69 |
| Design in [6] | 15 | 0.007 | 0.5 | 0.052 |
| Design in [7] | 35 | 0.02 | 0.5 | 0.35 |
| Design in [22] | 26 | 0.03 | 0.5 | 0.39 |
| Design in [8] | 19 | 0.012 | 0.5 | 0.114 |
| Design in [23] | 18 | 0.01 | 0.5 | 0.09 |
| Design in [24] | 14 | 0.01 | 0.5 | 0.07 |
| Our proposed (First approach) | 14 | 0.01 | 0.5 | 0.07 |
| Our proposed (Second approach) | 12 | 0.01 | 0.5 | 0.06 |

**5.2 The power dissipation analysis of the proposed XOR and Full-Adder**

We used the QCAPro tool to estimate the energy dissipation of our QCA designs. The QCAPro software uses approximations to calculate energy dissipation. Total energy dissipation is the sum of leakage and switching dissipation. The QCAPro software calculates the energy consumption of QCA circuits [25]. In our simulations, we have used three values of tunneling energy (0.5, 1, and $1.5E_k$). We present the power dissipation maps for the proposed XOR at 2K temperature in figures 11 and 12. A darker cell has more thermal hotspots dissipating heat.

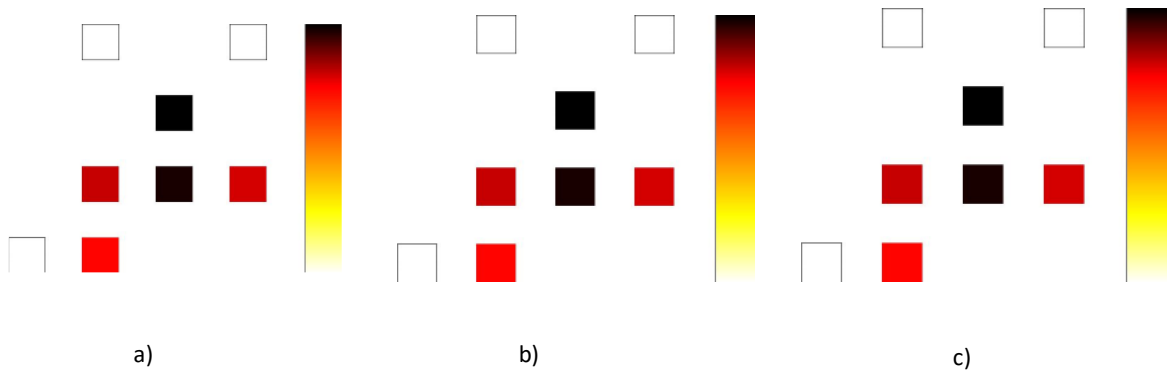

Figure 11. The Power dissipation maps (a: $0.5E_k$; b: $1E_k$; c: $1.5E_k$) of the proposed QCA XOR

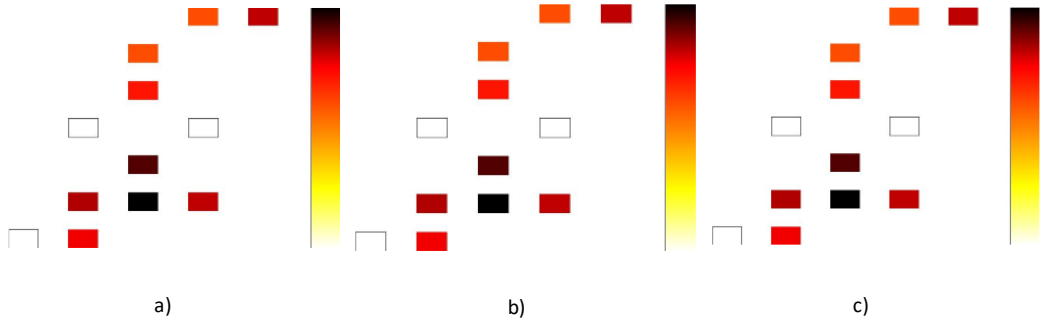

Figure 12. The Power dissipation maps (a: $0.5E_k$; b: $1E_k$; c: $1.5E_k$) of the proposed QCA Full-Adder

Tables 9 and 10 compare the energy dissipation estimation of the proposed XOR and Full-adder in QCA technology with the previous works. Note that some of the existing works didn't analyze energy dissipation. To the best of our knowledge, all the papers that reported energy dissipation are considered and compared with our works.

Table 9. The Energy dissipation results of different QCA XORs

| Designs | Avg. leakage energy diss. (eV) | | | Avg. switching energy diss. (eV) | | | Total Energy diss. (eV) | | |
|---|---|---|---|---|---|---|---|---|---|
| | $0.5\,E_k$ | $1.0\,E_k$ | $1.5\,E_k$ | $0.5\,E_k$ | $1.0\,E_k$ | $1.5\,E_k$ | $0.5\,E_k$ | $1.0\,E_k$ | $1.5\,E_k$ |
| [26] | 0.011 | 0.031 | 0.054 | 0.035 | 0.030 | 0.036 | 0.047 | 0.062 | 0.080 |
| [11] | 0.019 | 0.059 | 0.106 | 0.126 | 0.112 | 0.097 | 0.146 | 0.171 | 0.204 |
| [14] | 0.009 | 0.029 | 0.052 | 0.040 | 0.033 | 0.028 | 0.049 | 0.063 | 0.080 |
| [27] | 0.009 | 0.027 | 0.047 | 0.030 | 0.035 | 0.040 | 0.050 | 0.062 | 0.077 |
| [7] | 0.007 | 0.017 | 0.028 | 0.008 | 0.006 | 0.005 | 0.015 | 0.024 | 0.033 |
| Proposed | 0.002 | 0.006 | 0.010 | 0.002 | 0.002 | 0.001 | 0.005 | 0.008 | 0.012 |

Table 10. The Energy dissipation results of different QCA Full-Adders

| Designs | Avg. leakage energy diss. (eV) | | | Avg. switching energy diss. (eV) | | | Total Energy diss. (eV) | | |
|---|---|---|---|---|---|---|---|---|---|
| | $0.5\,E_k$ | $1.0\,E_k$ | $1.5\,E_k$ | $0.5\,E_k$ | $1.0\,E_k$ | $1.5\,E_k$ | $0.5\,E_k$ | $1.0\,E_k$ | $1.5\,E_k$ |
| [28] | 0.014 | 0.039 | 0.068 | 0.050 | 0.044 | 0.038 | 0.064 | 0.083 | 0.106 |
| [29] | 0.013 | 0.032 | 0.051 | 0.020 | 0.017 | 0.014 | 0.034 | 0.049 | 0.065 |
| [30] | 0.015 | 0.050 | 0.091 | 0.102 | 0.090 | 0.077 | 0.118 | 0.140 | 0.169 |
| [11] | 0.026 | 0.081 | 0.147 | 0.159 | 0.140 | 0.121 | 0.185 | 0.221 | 0.269 |
| [2] | 0.0082 | 0.020 | 0.033 | 0.014 | 0.011 | 0.0089 | 0.022 | 0.0319 | 0.0428 |
| [19] | 0.018 | 0.046 | 0.077 | 0.037 | 0.031 | 0.026 | 0.048 | 0.077 | 0.103 |
| Proposed | 0.0052 | 0.0121 | 0.0190 | 0.0037 | 0.0026 | 0.0020 | 0.0089 | 0.0147 | 0.021 |

## 6. Conclusion

The XOR gate is one of the fundamental gates utilized in many logic circuits. An efficient XOR gate in QCA can vastly improve efficiency in computational circuits. We suggested two novel structures for 3-inputs QCA XOR. Then we designed two new QCA full adders. The second proposed XOR has only 8 QCA cells. To the best of our knowledge, the proposed design exceeds the properties of the state-of-the-art 3-input QCA XORs and QCA full adders. The second suggested QCA full adder has only 12 QCA cells. It uses only normal cells. The suggested QCA XORs and full adders were proposed and simulated in a single layer. We used QCA Pro for calculating the energy of the proposed QCA XORs and full-adders. The result of the suggested structures was verified using simulations. One of the most important specifications of our designs is their scalability. Proposed QCA XORs and Full Adders are scalable, and they seem to provide one with a feasible alternative to the existing structures.

## Declarations

**Competing interests**: The authors declare that they have no known competing financial interests or personal relationships that could have appeared to influence the work reported in this paper.

**Authors' contributions:** All authors contributed equally to the study's conception and design. All authors read and approved the final manuscript.

**Availability of data and materials**: All relevant data are included in the article and the appendix.

## References

[1]  B. Safaiezadeh, E. Mahdipour, M. Haghparas, S. Sayedsalehi and M. Hosseinzadeh, "Novel design and simulation of reversible ALU in quantum dot cellular automata," *The Journal of Supercomputing,* pp. 868-882, 2021.

[2]  S.-S. Ahmadpour, M. Mosleh and S. Rasouli Heikalabad, "A revolution in nanostructure designs by proposing a novel QCA full-adder based on optimized 3-input XOR," *Physica B: Condensed Matter ,* pp. 383-392, 2018.

[3]  A. H. Majeed, M. S. B. Zainal, E. Alkaldy and D. M. Nor, "Full adder circuit design with novel lower complexity XOR gate in QCA technology," *Transactions on Electrical and Electronic Materials ,* pp. 198-207, 2020.

[4]  A. H. Majeed, M. S. Zainal, E. Alkaldy and D. M. Nor, "Single-bit comparator in quantum-dot cellular automata (QCA) technology using novel QCA-XNOR gates," *Journal of Electronic Science and Technology,* p. 100078, 2021.

[5]  S. S. and N. Navimipour, "A Three levels line-based full adder designing based on nano scale quantum-dot cellular automata," *Optik-International Journal for Light and Electron Optics,* 2017.

[6]  A. Majeed and E. Alkaldy, "High-performance adder using a new XOR gate in QCA,"*,* The journal of Supercomputing, pp. 11564-11579, 2022.

[7]  N. Safoev and J. Jun-Cheol , "A novel controllable inverter and adder/subtractor in quantum-dot cellular automata using cell interaction based XOR gate," *Microelectronic Engineering,* p. 111197, 2020.

[8]  F. Salimzadeh and S. Rasouli Heikalabad, "A full adder structure with a unique XNOR gate based on Coulomb interaction in QCA nanotechnology," *Optical and Quantum Electronics,* pp. 1-9, 2021.

[9]  L. C. McDermott, "Research on conceptual understanding in mechanics," *Physics today,* pp. 24-32, 1984.

[10] D. Bahrepour and N. Maroufi, "A 2-bit Full Comparator Design with Minimum Quantum Cost Function in Quantum-Dot Cellular Automata," *Journal of Information Systems and Telecommunication (JIST),* pp. 197-203, 2019.

[11] S. Angizi, E. Alkaldy, N. Bagherzadeh and K. Navi, "Novel robust single layer wire crossing approach for exclusive or sum of products logic design with quantum-dot cellular automata," *Journal of Low Power Electronics,* pp. 259-271, 2014.

[12] S. Shadi, S. Angizi, M. H. Moaiyeri and S. Sayedsalehi, "Designing efficient QCA logical circuits with power dissipation analysis," *Microelectronics Journal,* pp. 462-471, 2015.

[13] H. Dallaki and M. Mehran, "Novel subtractor design based on quantum-dot cellular automata (QCA) nanotechnology," *International Journal of Nanoscience and Nanotechnology,* pp. 257-262, 2015.

[14] G. Singh, R. K. Sarin and B. Raj, "A novel robust exclusive-OR function implementation in QCA nanotechnology with energy dissipation analysis," *Journal of Computational Electronics,* pp. 455-465, 2016.

[15] A. N. Bahar, S. Waheed, N. Hossain and M. Asaduzzaman, "A novel 3-input XOR function implementation in quantum dot-cellular automata with energy dissipation analysis," *Alexandria Engineering Journal,* pp. 729-738, 2018.

[16] R. H.R. and A. Rezai, "Novel efficient circuit design for multilayer QCA RCA," *Int. J. theor. Phys.,* pp. 1745-1757, 2019.

[17] S. Seyedi and N. Jafari Navimipour, "An optimized design of full adder based on nanoscale quantum-dot cellular automata," *Optik ,* pp. 243-256, 2018.

[18] S. Sarmadi, S. Sayedsalehi, M. Fartash and S. Angizi, "A structured ultra-dense QCA one-bit full-adder cell," *Quantum Matter,* pp. 118-123, 2016.

[19] S. Zoka and M. Gholami, "A novel efficient full adder–subtractor in QCA nanotechnology.," *International Nano Letters ,* pp. 51-54, 2019.

[20] M. Patidar, A. Shrivastava, S. Miah, Y. Kumar and A. K. Sivaraman, "An energy efficient high-speed quantum-dot based full adder design and parity gate for nano application," *Materials Today,* 2022.

[21] M. Zahmatkesh, S. Tabrizchi, S. Mohammadyan, K. Navi and N. Bagherzadeh, "Robust coplanar full adder based on novel inverter in quantum cellular automata," *International Journal of Theoretical Physics,* pp. 639-655, 2019.

[22] S. Babaie, A. Sadoghifar and A. Newaz Bahar, "Design of an efficient multilayer arithmetic logic unit in quantum-dot cellular automata (QCA)," *IEEE Transactions on Circuits and Systems II: Express Briefs,* pp. 963-967, 2018.

[23] M. M. Abutaleb, "Utilizing charge reconfigurations of quantum-dot cells in building blocks to design nanoelectronic adder circuits," *Computers & Electrical Engineering,* p. 106712, 2020.

[24] I. Gassoumi, L. Touil and A. Mtibaa, "An efficient QCA-Based full adder design with Power dissipation analysis," *International Journal of Electronics Letters ,* pp. 1-13, 2022.

[25] S. Srivastava, A. Asthana, S. Bhanja and S. Sarkar, "QCAPro-an error-power estimation tool for QCA circuit design," *2011 IEEE international symposium of circuits and systems,* pp. 2377-2380, 2011.

[26] S. Sheikhfaal, S. Angizi, S. Sarmadi, M. H. Moaiyeri and S. Sayedsalehi, "Designing efficient QCA logical circuits with power dissipation analysis," *Microelectronics Journal,* pp. 462-471, 2015.

[27] A. M. Chabi, S. Sayedsalehi, S. Angizi and K. Navi, "Efficient QCA exclusive-or and multiplexer circuits based on a nanoelectronic-compatible designing approach," *International scholarly research notices,* 2014.

[28] S. R. Heikalabad, M. Naji Asfestani and M. Hosseinzadeh, "A full adder structure without cross-wiring in quantum-dot cellular automata with energy dissipation analysis," *The Journal of Supercomputing,* pp. 1994-2005, 2018.

[29] M. Balali, A. Rezai, H. Balali, F. Rabiei and S. Emadi, "Towards coplanar quantum-dot cellular automata adders based on efficient three-input XOR gate," *Results in physics,* pp. 1389-1395, 2017.

[30] D. Abedi, G. Jaberipur and M. Sangsefidi, "Coplanar full adder in quantum-dot cellular automata via clock-zone-based crossover," *IEEE transactions on nanotechnology,* pp. 497-504, 2015.

# Appendix

We provide the simulation results for kink energy between the output cell electrons and other cells' electrons in the proposed structures. Table I represents the values calculated when the output of the structure is equal to one. For the electrons at position X, we calculated the kink energy affected by all electrons. The right column shows the amount of kink energy caused by electrons on the electron at position Y.

Table I. Calculation of kink energy between output cell electrons and all electrons in Fig. 4 b.

| For electron X | For electron Y |
|---|---|
| $U_{1X} = \frac{A}{r_{1X}} = \frac{23.04 * 10^{-29}}{45.65 * 10^{-9}} = 0.51 * 10^{-20}$ | $U_{1Y} = \frac{A}{r_{1Y}} = \frac{23.04 * 10^{-29}}{69.32 * 10^{-9}} = 0.33 * 10^{-20}$ |
| $U_{2X} = \frac{A}{r_{2X}} = \frac{23.04 * 10^{-29}}{44.72 * 10^{-9}} = 0.52 * 10^{-20}$ | $U_{2Y} = \frac{A}{r_{2Y}} = \frac{23.04 * 10^{-29}}{69.33 * 10^{-9}} = 0.33 * 10^{-20}$ |
| $U_{3X} = \frac{A}{r_{3X}} = \frac{23.04 * 10^{-29}}{40 * 10^{-9}} = 0.58 * 10^{-20}$ | $U_{3Y} = \frac{A}{r_{3Y}} = \frac{23.04 * 10^{-29}}{60.72 * 10^{-9}} = 0.38 * 10^{-20}$ |
| $U_{4X} = \frac{A}{r_{4X}} = \frac{23.04 * 10^{-29}}{29.73 * 10^{-9}} = 0.77 * 10^{-20}$ | $U_{4Y} = \frac{A}{r_{4Y}} = \frac{23.04 * 10^{-29}}{38 * 10^{-9}} = 0.61 * 10^{-20}$ |
| $U_{5X} = \frac{A}{r_{5X}} = \frac{23.04 * 10^{-29}}{20.09 * 10^{-9}} = 1.15 * 10^{-20}$ | $U_{5Y} = \frac{A}{r_{5Y}} = \frac{23.04 * 10^{-29}}{42.94 * 10^{-9}} = 0.54 * 10^{-20}$ |
| $U_{6X} = \frac{A}{r_{6X}} = \frac{23.04 * 10^{-29}}{20.09 * 10^{-9}} = 1.15 * 10^{-20}$ | $U_{6Y} = \frac{A}{r_{6Y}} = \frac{23.04 * 10^{-29}}{42.94 * 10^{-9}} = 0.54 * 10^{-20}$ |
| $U_{7X} = \frac{A}{r_{7X}} = \frac{23.04 * 10^{-29}}{22 * 10^{-9}} = 1.05 * 10^{-20}$ | $U_{7Y} = \frac{A}{r_{7Y}} = \frac{23.04 * 10^{-29}}{42.94 * 10^{-9}} = 0.54 * 10^{-20}$ |
| $U_{8X} = \frac{A}{r_{8X}} = \frac{23.04 * 10^{-29}}{26.90 * 10^{-9}} = 0.86 * 10^{-20}$ | $U_{8Y} = \frac{A}{r_{8Y}} = \frac{23.04 * 10^{-29}}{58 * 10^{-9}} = 0.38 * 10^{-20}$ |
| $U_{9X} = \frac{A}{r_{9X}} = \frac{23.04 * 10^{-29}}{2 * 10^{-9}} = 11.52 * 10^{-20}$ | $U_{9Y} = \frac{A}{r_{9Y}} = \frac{23.04 * 10^{-29}}{26.90 * 10^{-9}} = 0.86 * 10^{-20}$ |
| $U_{10X} = \frac{A}{r_{10X}} = \frac{23.04 * 10^{-29}}{26.90 * 10^{-9}} = 0.86 * 10^{-20}$ | $U_{10Y} = \frac{A}{r_{10Y}} = \frac{23.04 * 10^{-29}}{38 * 10^{-9}} = 0.61 * 10^{-20}$ |
| $U_{11X} = \frac{A}{r_{11X}} = \frac{23.04 * 10^{-29}}{63.24 * 10^{-9}} = 0.36 * 10^{-20}$ | $U_{11Y} = \frac{A}{r_{11Y}} = \frac{23.04 * 10^{-29}}{60.03 * 10^{-9}} = 0.38 * 10^{-20}$ |
| $U_{12X} = \frac{A}{r_{12X}} = \frac{23.04 * 10^{-29}}{68.20 * 10^{-9}} = 0.34 * 10^{-20}$ | $U_{12Y} = \frac{A}{r_{12Y}} = \frac{23.04 * 10^{-29}}{80.05 * 10^{-9}} = 0.29 * 10^{-20}$ |
| $U_{13X} = \frac{A}{r_{13X}} = \frac{23.04 * 10^{-29}}{44.72 * 10^{-9}} = 0.52 * 10^{-20}$ | $U_{13Y} = \frac{A}{r_{13Y}} = \frac{23.04 * 10^{-29}}{58.03 * 10^{-9}} = 0.38 * 10^{-20}$ |
| $U_{14X} = \frac{A}{r_{14X}} = \frac{23.04 * 10^{-29}}{58.06 * 10^{-9}} = 0.40 * 10^{-20}$ | $U_{14Y} = \frac{A}{r_{14Y}} = \frac{23.04 * 10^{-29}}{44.72 * 10^{-9}} = 0.52 * 10^{-20}$ |

Table II shows the kink energy between electrons $X_1$ and $Y_1$ of the output cell with other electrons. Table II assumes the output of Carry of the proposed full adder is one. We calculated the kink energy of all electrons on the electron at position $X_1$ in the left column. The right column shows the amount of the kink energy affected by all electrons on the electron at position $Y_1$.

Table II. Calculation of Kink energy between all electrons and $X_1$ and $Y_1$ electrons of the output cell of Fig. 9

| For electron $X_1$ | For electron $Y_1$ |
|---|---|
| $U_{1X_1} = \dfrac{A}{r_{1X}} = \dfrac{23.04 * 10^{-29}}{20 * 10^{-9}} = 1.152 * 10^{-20}$ | $U_{1Y_1} = \dfrac{A}{r_{1Y}} = \dfrac{23.04 * 10^{-29}}{18.11 * 10^{-9}} = 1.272 * 10^{-20}$ |
| $U_{2X_1} = \dfrac{A}{r_{2X}} = \dfrac{23.04 * 10^{-29}}{42.05 * 10^{-9}} = 0.547 * 10^{-20}$ | $U_{2Y_1} = \dfrac{A}{r_{2Y}} = \dfrac{23.04 * 10^{-29}}{20 * 10^{-9}} = 1.152 * 10^{-20}$ |
| $U_{3X_1} = \dfrac{A}{r_{3X}} = \dfrac{23.04 * 10^{-29}}{61.35 * 10^{-9}} = 0.375 * 10^{-20}$ | $U_{3Y_1} = \dfrac{A}{r_{3Y}} = \dfrac{23.04 * 10^{-29}}{40.04 * 10^{-9}} = 0.575 * 10^{-20}$ |
| $U_{4X_1} = \dfrac{A}{r_{4X}} = \dfrac{23.04 * 10^{-29}}{55.17 * 10^{-9}} = 0.417 * 10^{-20}$ | $U_{4Y_1} = \dfrac{A}{r_{4Y}} = \dfrac{23.04 * 10^{-29}}{29.73 * 10^{-9}} = 0.774 * 10^{-20}$ |
| $U_{5X_1} = \dfrac{A}{r_{5X}} = \dfrac{23.04 * 10^{-29}}{70.45 * 10^{-9}} = 0.327 * 10^{-20}$ | $U_{5Y_1} = \dfrac{A}{r_{5Y}} = \dfrac{23.04 * 10^{-29}}{47.41 * 10^{-9}} = 0.485 * 10^{-20}$ |
| $U_{6X_1} = \dfrac{A}{r_{6X}} = \dfrac{23.04 * 10^{-29}}{70.45 * 10^{-9}} = 0.327 * 10^{-20}$ | $U_{6Y_1} = \dfrac{A}{r_{6Y}} = \dfrac{23.04 * 10^{-29}}{45.65 * 10^{-9}} = 0.504 * 10^{-20}$ |
| $U_{7X_1} = \dfrac{A}{r_{7X}} = \dfrac{23.04 * 10^{-29}}{84.85 * 10^{-9}} = 0.271 * 10^{-20}$ | $U_{7Y_1} = \dfrac{A}{r_{7Y}} = \dfrac{23.04 * 10^{-29}}{59.39 * 10^{-9}} = 0.387 * 10^{-20}$ |
| $U_{8X_1} = \dfrac{A}{r_{8X}} = \dfrac{23.04 * 10^{-29}}{110.3 * 10^{-9}} = 0.208 * 10^{-20}$ | $U_{8Y_1} = \dfrac{A}{r_{8Y}} = \dfrac{23.04 * 10^{-29}}{84.85 * 10^{-9}} = 0.271 * 10^{-20}$ |
| $U_{9X_1} = \dfrac{A}{r_{9X}} = \dfrac{23.04 * 10^{-29}}{61.35 * 10^{-9}} = 0.375 * 10^{-20}$ | $U_{9Y_1} = \dfrac{A}{r_{9Y}} = \dfrac{23.04 * 10^{-29}}{42.05 * 10^{-9}} = 0.547 * 10^{-20}$ |
| $U_{10X_1} = \dfrac{A}{r_{10X}} = \dfrac{23.04 * 10^{-29}}{86.76 * 10^{-9}} = 0.265 * 10^{-20}$ | $U_{10Y_1} = \dfrac{A}{r_{10Y}} = \dfrac{23.04 * 10^{-29}}{63.24 * 10^{-9}} = 0.364 * 10^{-20}$ |
| $U_{11X_1} = \dfrac{A}{r_{11X}} = \dfrac{23.04 * 10^{-29}}{98.81 * 10^{-9}} = 0.233 * 10^{-20}$ | $U_{11Y_1} = \dfrac{A}{r_{11Y}} = \dfrac{23.04 * 10^{-29}}{73.78 * 10^{-9}} = 0.312 * 10^{-20}$ |
| $U_{12X_1} = \dfrac{A}{r_{12X}} = \dfrac{23.04 * 10^{-29}}{105.8 * 10^{-9}} = 0.217 * 10^{-20}$ | $U_{12Y_1} = \dfrac{A}{r_{12Y}} = \dfrac{23.04 * 10^{-29}}{83 * 10^{-9}} = 0.277 * 10^{-20}$ |
| $U_{13X_1} = \dfrac{A}{r_{13X}} = \dfrac{23.04 * 10^{-29}}{126.8 * 10^{-9}} = 0.181 * 10^{-20}$ | $U_{13Y_1} = \dfrac{A}{r_{13Y}} = \dfrac{23.04 * 10^{-29}}{101.6 * 10^{-9}} = 0.226 * 10^{-20}$ |
| $U_{14X_1} = \dfrac{A}{r_{14X}} = \dfrac{23.04 * 10^{-29}}{132.4 * 10^{-9}} = 0.174 * 10^{-20}$ | $U_{14Y_1} = \dfrac{A}{r_{14Y}} = \dfrac{23.04 * 10^{-29}}{108.46 * 10^{-9}} = 0.212 * 10^{-20}$ |
| $U_{15X_1} = \dfrac{A}{r_{15X}} = \dfrac{23.04 * 10^{-29}}{115.6 * 10^{-9}} = 0.199 * 10^{-20}$ | $U_{15Y_1} = \dfrac{A}{r_{15Y}} = \dfrac{23.04 * 10^{-29}}{91.23 * 10^{-9}} = 0.252 * 10^{-20}$ |
| $U_{16X_1} = \dfrac{A}{r_{16X}} = \dfrac{23.04 * 10^{-29}}{124.6 * 10^{-9}} = 0.184 * 10^{-20}$ | $U_{16Y_1} = \dfrac{A}{r_{16Y}} = \dfrac{23.04 * 10^{-29}}{102.39 * 10^{-9}} = 0.225 * 10^{-20}$ |
| $U_{17X_1} = \dfrac{A}{r_{17X}} = \dfrac{23.04 * 10^{-29}}{156.2 * 10^{-9}} = 0.147 * 10^{-20}$ | $U_{17Y_1} = \dfrac{A}{r_{17Y}} = \dfrac{23.04 * 10^{-29}}{129.6 * 10^{-9}} = 0.177 * 10^{-20}$ |
| $U_{18X_1} = \dfrac{A}{r_{18X}} = \dfrac{23.04 * 10^{-29}}{159.5 * 10^{-9}} = 0.144 * 10^{-20}$ | $U_{18Y_1} = \dfrac{A}{r_{18Y}} = \dfrac{23.04 * 10^{-29}}{130.1 * 10^{-9}} = 0.176 * 10^{-20}$ |
| $U_{19X_1} = \dfrac{A}{r_{19X}} = \dfrac{23.04 * 10^{-29}}{143.1 * 10^{-9}} = 0.160 * 10^{-20}$ | $U_{19Y_1} = \dfrac{A}{r_{19Y}} = \dfrac{23.04 * 10^{-29}}{116.59 * 10^{-9}} = 0.197 * 10^{-20}$ |
| $U_{20X_1} = \dfrac{A}{r_{20X}} = \dfrac{23.04 * 10^{-29}}{150.5 * 10^{-9}} = 0.153 * 10^{-20}$ | $U_{20Y_1} = \dfrac{A}{r_{20Y}} = \dfrac{23.04 * 10^{-29}}{127.13 * 10^{-9}} = 0.181 * 10^{-20}$ |

Table III shows the kink energy between electrons $X_2$ and $Y_2$ of the output cell with other electrons. We assume that the output of Sum of the proposed full adder is one. The left column shows the kink energy of all electrons on the electron at position $X_2$. On the right column, we see how much kink energy is affected by all electrons on the electron at position $Y_2$.

Table III. Calculation of the Kink energy between all electrons with $X_2$ and $Y_2$ electrons of the output cell of Fig. 9

| For electron $X_2$ | For electron $Y_2$ |
|---|---|
| $U_{1X_2} = \dfrac{A}{r_1} = \dfrac{23.04 * 10^{-29}}{98 * 10^{-9}} = 0.235 * 10^{-20}$ | $U_{1Y_2} = \dfrac{A}{r_1} = \dfrac{23.04 * 10^{-29}}{119.36 * 10^{-9}} = 0.193 * 10^{-20}$ |
| $U_{2X_2} = \dfrac{A}{r_{2X}} = \dfrac{23.04 * 10^{-29}}{83.95 * 10^{-9}} = 0.274 * 10^{-20}$ | $U_{2Y_2} = \dfrac{A}{r_{2Y}} = \dfrac{23.04 * 10^{-29}}{100 * 10^{-9}} = 0.230 * 10^{-20}$ |
| $U_{3X_2} = \dfrac{A}{r_{3X}} = \dfrac{23.04 * 10^{-29}}{88.56 * 10^{-9}} = 0.260 * 10^{-20}$ | $U_{3Y_2} = \dfrac{A}{r_{3Y}} = \dfrac{23.04 * 10^{-29}}{82.46 * 10^{-9}} = 0.279 * 10^{-20}$ |
| $U_{4X_2} = \dfrac{A}{r_{4X}} = \dfrac{23.04 * 10^{-29}}{65.14 * 10^{-9}} = 0.353 * 10^{-20}$ | $U_{4Y_2} = \dfrac{A}{r_{4Y}} = \dfrac{23.04 * 10^{-29}}{78.02 * 10^{-9}} = 0.290 * 10^{-20}$ |
| $U_{5X_2} = \dfrac{A}{r_{5X}} = \dfrac{23.04 * 10^{-29}}{71.02 * 10^{-9}} = 0.324 * 10^{-20}$ | $U_{5Y_2} = \dfrac{A}{r_{5Y}} = \dfrac{23.04 * 10^{-29}}{80.52 * 10^{-9}} = 0.286 * 10^{-20}$ |
| $U_{6X_2} = \dfrac{A}{r_{6X}} = \dfrac{23.04 * 10^{-29}}{44.72 * 10^{-9}} = 0.515 * 10^{-20}$ | $U_{6Y_2} = \dfrac{A}{r_{6Y}} = \dfrac{23.04 * 10^{-29}}{58.03 * 10^{-9}} = 0.397 * 10^{-20}$ |
| $U_{7X_2} = \dfrac{A}{r_{7X}} = \dfrac{23.04 * 10^{-29}}{56.56 * 10^{-9}} = 0.407 * 10^{-20}$ | $U_{7Y_2} = \dfrac{A}{r_{7Y}} = \dfrac{23.04 * 10^{-29}}{62.03 * 10^{-9}} = 0.371 * 10^{-20}$ |
| $U_{8X_2} = \dfrac{A}{r_{8X}} = \dfrac{23.04 * 10^{-29}}{62.03 * 10^{-9}} = 0.371 * 10^{-20}$ | $U_{8Y_2} = \dfrac{A}{r_{8Y}} = \dfrac{23.04 * 10^{-29}}{56.56 * 10^{-9}} = 0.407 * 10^{-20}$ |
| $U_{9X_2} = \dfrac{A}{r_{9X}} = \dfrac{23.04 * 10^{-29}}{40 * 10^{-9}} = 0.576 * 10^{-20}$ | $U_{9Y_2} = \dfrac{A}{r_{9Y}} = \dfrac{23.04 * 10^{-29}}{60.72 * 10^{-9}} = 0.379 * 10^{-20}$ |
| $U_{10X_2} = \dfrac{A}{r_{10X}} = \dfrac{23.04 * 10^{-29}}{26.9 * 10^{-9}} = 0.856 * 10^{-20}$ | $U_{10Y_2} = \dfrac{A}{r_{10Y}} = \dfrac{23.04 * 10^{-29}}{40 * 10^{-9}} = 0.576 * 10^{-20}$ |
| $U_{11X_2} = \dfrac{A}{r_{11X}} = \dfrac{23.04 * 10^{-29}}{42.94 * 10^{-9}} = 0.536 * 10^{-20}$ | $U_{11Y_2} = \dfrac{A}{r_{11Y}} = \dfrac{23.04 * 10^{-29}}{42.94 * 10^{-9}} = 0.536 * 10^{-20}$ |
| $U_{12X_2} = \dfrac{A}{r_{12X}} = \dfrac{23.04 * 10^{-29}}{20.09 * 10^{-9}} = 1.146 * 10^{-20}$ | $U_{12Y_2} = \dfrac{A}{r_{12Y}} = \dfrac{23.04 * 10^{-29}}{20.09 * 10^{-9}} = 1.146 * 10^{-20}$ |
| $U_{13X_2} = \dfrac{A}{r_{13X}} = \dfrac{23.04 * 10^{-29}}{58 * 10^{-9}} = 0.397 * 10^{-20}$ | $U_{13Y_2} = \dfrac{A}{r_{13Y}} = \dfrac{23.04 * 10^{-29}}{43.86 * 10^{-9}} = 0.525 * 10^{-20}$ |
| $U_{14X_2} = \dfrac{A}{r_{14X}} = \dfrac{23.04 * 10^{-29}}{43.86 * 10^{-9}} = 0.526 * 10^{-20}$ | $U_{14Y_2} = \dfrac{A}{r_{14Y}} = \dfrac{23.04 * 10^{-29}}{22 * 10^{-9}} = 1.047 * 10^{-20}$ |
| $U_{15X_2} = \dfrac{A}{r_{15X}} = \dfrac{23.04 * 10^{-29}}{38 * 10^{-9}} = 0.606 * 10^{-20}$ | $U_{15Y_2} = \dfrac{A}{r_{15Y}} = \dfrac{23.04 * 10^{-29}}{26.9 * 10^{-9}} = 0.856 * 10^{-20}$ |
| $U_{16X_2} = \dfrac{A}{r_{16X}} = \dfrac{23.04 * 10^{-29}}{26.9 * 10^{-9}} = 0.856 * 10^{-20}$ | $U_{16Y_2} = \dfrac{A}{r_{16Y}} = \dfrac{23.04 * 10^{-29}}{2 * 10^{-9}} = 11.52 * 10^{-20}$ |
| $U_{17X_2} = \dfrac{A}{r_{17X}} = \dfrac{23.04 * 10^{-29}}{80.52 * 10^{-9}} = 0.286 * 10^{-20}$ | $U_{17Y_2} = \dfrac{A}{r_{17Y}} = \dfrac{23.04 * 10^{-29}}{60.03 * 10^{-9}} = 0.383 * 10^{-20}$ |
| $U_{18X_2} = \dfrac{A}{r_{18X}} = \dfrac{23.04 * 10^{-29}}{71.02 * 10^{-9}} = 0.324 * 10^{-20}$ | $U_{18Y_2} = \dfrac{A}{r_{18Y}} = \dfrac{23.04 * 10^{-29}}{46.51 * 10^{-9}} = 0.495 * 10^{-20}$ |
| $U_{19X_2} = \dfrac{A}{r_{19X}} = \dfrac{23.04 * 10^{-29}}{84.42 * 10^{-9}} = 0.272 * 10^{-20}$ | $U_{19Y_2} = \dfrac{A}{r_{19Y}} = \dfrac{23.04 * 10^{-29}}{40.04 * 10^{-9}} = 0.575 * 10^{-20}$ |
| $U_{20X_2} = \dfrac{A}{r_{20X}} = \dfrac{23.04 * 10^{-29}}{55.17 * 10^{-9}} = 0.417 * 10^{-20}$ | $U_{20Y_2} = \dfrac{A}{r_{20Y}} = \dfrac{23.04 * 10^{-29}}{29.73 * 10^{-9}} = 0.774 * 10^{-20}$ |

The kink energy between electrons $X_3$ and $Y_3$ of the output cell is shown in Table IV. The values in Table IV are calculated assuming that the structure's output or Carry is zero. Our left column shows the amount of kink energy influenced by all electrons on the electron at the position $X_3$. Our right column shows how much kink energy the electron at the position $Y_3$ is affected by all electrons.

Table IV. Calculation of the Kink energy between all electrons with $X_3$ and $Y_3$ electrons of the output cell of Fig. 9

| For electron $X_3$ | For electron $Y_3$ |
|---|---|
| $U_{1X_3} = \frac{A}{r_1} = \frac{23.04 * 10^{-29}}{2 * 10^{-9}} = 11.52 * 10^{-20}$ | $U_{1Y_3} = \frac{A}{r_{1Y}} = \frac{23.04 * 10^{-29}}{26.90 * 10^{-9}} = 0.856 * 10^{-20}$ |
| $U_{2X_3} = \frac{A}{r_{2X}} = \frac{23.04 * 10^{-29}}{26.90 * 10^{-9}} = 0.856 * 10^{-20}$ | $U_{2Y_3} = \frac{A}{r_{2Y}} = \frac{23.04 * 10^{-29}}{38 * 10^{-9}} = 0.606 * 10^{-20}$ |
| $U_{3X_3} = \frac{A}{r_{3X}} = \frac{23.04 * 10^{-29}}{44.72 * 10^{-9}} = 0.515 * 10^{-20}$ | $U_{3Y_3} = \frac{A}{r_{3Y}} = \frac{23.04 * 10^{-29}}{58.03 * 10^{-9}} = 0.397 * 10^{-20}$ |
| $U_{4X_3} = \frac{A}{r_{4X}} = \frac{23.04 * 10^{-29}}{43.9 * 10^{-9}} = 0.524 * 10^{-20}$ | $U_{4Y_3} = \frac{A}{r_{4Y}} = \frac{23.04 * 10^{-29}}{44.72 * 10^{-9}} = 0.515 * 10^{-20}$ |
| $U_{5X_3} = \frac{A}{r_{5X}} = \frac{23.04 * 10^{-29}}{56.56 * 10^{-9}} = 0.407 * 10^{-20}$ | $U_{5Y_3} = \frac{A}{r_{5Y}} = \frac{23.04 * 10^{-29}}{62.03 * 10^{-9}} = 0.371 * 10^{-20}$ |
| $U_{6X_3} = \frac{A}{r_{6X}} = \frac{23.04 * 10^{-29}}{70.45 * 10^{-9}} = 0.327 * 10^{-20}$ | $U_{6Y_3} = \frac{A}{r_{6Y}} = \frac{23.04 * 10^{-29}}{56.56 * 10^{-9}} = 0.407 * 10^{-20}$ |
| $U_{7X_3} = \frac{A}{r_{7X}} = \frac{23.04 * 10^{-29}}{73.23 * 10^{-9}} = 0.314 * 10^{-20}$ | $U_{7Y_3} = \frac{A}{r_{7Y}} = \frac{23.04 * 10^{-29}}{73.23 * 10^{-9}} = 0.314 * 10^{-20}$ |
| $U_{8X_3} = \frac{A}{r_{8X}} = \frac{23.04 * 10^{-29}}{86.27 * 10^{-9}} = 0.267 * 10^{-20}$ | $U_{8Y_3} = \frac{A}{r_{8Y}} = \frac{23.04 * 10^{-29}}{98.40 * 10^{-9}} = 0.234 * 10^{-20}$ |
| $U_{9X_3} = \frac{A}{r_{9X}} = \frac{23.04 * 10^{-29}}{60.03 * 10^{-9}} = 0.383 * 10^{-20}$ | $U_{9Y_3} = \frac{A}{r_{9Y}} = \frac{23.04 * 10^{-29}}{46.51 * 10^{-9}} = 0.495 * 10^{-20}$ |
| $U_{10X_3} = \frac{A}{r_{10X}} = \frac{23.04 * 10^{-29}}{78.02 * 10^{-9}} = 0.295 * 10^{-20}$ | $U_{10Y_3} = \frac{A}{r_{10Y}} = \frac{23.04 * 10^{-29}}{71.02 * 10^{-9}} = 0.324 * 10^{-20}$ |
| $U_{11X_3} = \frac{A}{r_{11X}} = \frac{23.04 * 10^{-29}}{90.35 * 10^{-9}} = 0.255 * 10^{-20}$ | $U_{11Y_3} = \frac{A}{r_{11Y}} = \frac{23.04 * 10^{-29}}{83.45 * 10^{-9}} = 0.276 * 10^{-20}$ |
| $U_{12X_3} = \frac{A}{r_{12X}} = \frac{23.04 * 10^{-29}}{100.43 * 10^{-9}} = 0.229 * 10^{-20}$ | $U_{12Y_3} = \frac{A}{r_{12Y}} = \frac{23.04 * 10^{-29}}{89.44 * 10^{-9}} = 0.257 * 10^{-20}$ |
| $U_{13X_3} = \frac{A}{r_{13X}} = \frac{23.04 * 10^{-29}}{116.6 * 10^{-9}} = 0.197 * 10^{-20}$ | $U_{13Y_3} = \frac{A}{r_{13Y}} = \frac{23.04 * 10^{-29}}{113.1 * 10^{-9}} = 0.203 * 10^{-20}$ |
| $U_{14X_3} = \frac{A}{r_{14X}} = \frac{23.04 * 10^{-29}}{125.2 * 10^{-9}} = 0.184 * 10^{-20}$ | $U_{14Y_3} = \frac{A}{r_{14Y}} = \frac{23.04 * 10^{-29}}{115.6 * 10^{-9}} = 0.199 * 10^{-20}$ |
| $U_{15X_3} = \frac{A}{r_{15X}} = \frac{23.04 * 10^{-29}}{108.46 * 10^{-9}} = 0.212 * 10^{-20}$ | $U_{15Y_3} = \frac{A}{r_{15Y}} = \frac{23.04 * 10^{-29}}{129.6 * 10^{-9}} = 0.177 * 10^{-20}$ |
| $U_{16X_3} = \frac{A}{r_{16X}} = \frac{23.04 * 10^{-29}}{120.03 * 10^{-9}} = 0.191 * 10^{-20}$ | $U_{16Y_3} = \frac{A}{r_{16Y}} = \frac{23.04 * 10^{-29}}{107.7 * 10^{-9}} = 0.213 * 10^{-20}$ |
| $U_{17X_3} = \frac{A}{r_{17X}} = \frac{23.04 * 10^{-29}}{144.2 * 10^{-9}} = 0.159 * 10^{-20}$ | $U_{17Y_3} = \frac{A}{r_{17Y}} = \frac{23.04 * 10^{-29}}{141.45 * 10^{-9}} = 0.162 * 10^{-20}$ |
| $U_{18X_3} = \frac{A}{r_{18X}} = \frac{23.04 * 10^{-29}}{153.1 * 10^{-9}} = 0.150 * 10^{-20}$ | $U_{18Y_3} = \frac{A}{r_{18Y}} = \frac{23.04 * 10^{-29}}{144.2 * 10^{-9}} = 0.159 * 10^{-20}$ |
| $U_{19X_3} = \frac{A}{r_{19X}} = \frac{23.04 * 10^{-29}}{134.1 * 10^{-9}} = 0.171 * 10^{-20}$ | $U_{19Y_3} = \frac{A}{r_{19Y}} = \frac{23.04 * 10^{-29}}{128.4 * 10^{-9}} = 0.179 * 10^{-20}$ |
| $U_{20X_3} = \frac{A}{r_{20X}} = \frac{23.04 * 10^{-29}}{144.1 * 10^{-9}} = 0.159 * 10^{-20}$ | $U_{20Y_3} = \frac{A}{r_{20Y}} = \frac{23.04 * 10^{-29}}{134.1 * 10^{-9}} = 0.171 * 10^{-20}$ |

Based on Table V, we determined the kink energy between electrons $X_4$ and $Y_4$ of the output cell with other electrons. Table V shows the results when the output of the structure or Sum is zero. The left column shows the amount of the kink energy affected by all electrons on the electron $X_4$. We calculated the kink energy of all electrons on the electron at the position $Y_4$ in the right column.

Table V. Calculation of the Kink energy between all electrons with $X_4$ and $Y_4$ electrons of the output cell of Fig. 9

| For electron $X_4$ | For electron $Y_4$ |
|---|---|
| $U_{1X_4} = \frac{A}{r_{1X}} = \frac{23.04 * 10^{-29}}{101.6 * 10^{-9}} = 0.226 * 10^{-20}$ | $U_{1Y_4} = \frac{A}{r_{1Y}} = \frac{23.04 * 10^{-29}}{118 * 10^{-9}} = 0.195 * 10^{-20}$ |
| $U_{2X_4} = \frac{A}{r_{2X}} = \frac{23.04 * 10^{-29}}{82 * 10^{-9}} = 0.280 * 10^{-20}$ | $U_{2Y_4} = \frac{A}{r_{2Y}} = \frac{23.04 * 10^{-29}}{101.6 * 10^{-9}} = 0.228 * 10^{-20}$ |
| $U_{3X_4} = \frac{A}{r_{3X}} = \frac{23.04 * 10^{-29}}{82.46 * 10^{-9}} = 0.279 * 10^{-20}$ | $U_{3Y_4} = \frac{A}{r_{3Y}} = \frac{23.04 * 10^{-29}}{105.1 * 10^{-9}} = 0.212 * 10^{-20}$ |
| $U_{4X_4} = \frac{A}{r_{4X}} = \frac{23.04 * 10^{-29}}{62.03 * 10^{-9}} = 0.371 * 10^{-20}$ | $U_{4Y_4} = \frac{A}{r_{4Y}} = \frac{23.04 * 10^{-29}}{82.46 * 10^{-9}} = 0.279 * 10^{-20}$ |
| $U_{5X_4} = \frac{A}{r_{5X}} = \frac{23.04 * 10^{-29}}{63.24 * 10^{-9}} = 0.364 * 10^{-20}$ | $U_{5Y_4} = \frac{A}{r_{5Y}} = \frac{23.04 * 10^{-29}}{86.76 * 10^{-9}} = 0.265 * 10^{-20}$ |
| $U_{6X_4} = \frac{A}{r_{6X}} = \frac{23.04 * 10^{-29}}{42.04 * 10^{-9}} = 0.548 * 10^{-20}$ | $U_{6Y_4} = \frac{A}{r_{6Y}} = \frac{23.04 * 10^{-29}}{63.90 * 10^{-9}} = 0.360 * 10^{-20}$ |
| $U_{7X_4} = \frac{A}{r_{7X}} = \frac{23.04 * 10^{-29}}{45.65 * 10^{-9}} = 0.504 * 10^{-20}$ | $U_{7Y_4} = \frac{A}{r_{7Y}} = \frac{23.04 * 10^{-29}}{70.45 * 10^{-9}} = 0.327 * 10^{-20}$ |
| $U_{8X_4} = \frac{A}{r_{8X}} = \frac{23.04 * 10^{-29}}{47.41 * 10^{-9}} = 0.485 * 10^{-20}$ | $U_{8Y_4} = \frac{A}{r_{8Y}} = \frac{23.04 * 10^{-29}}{70.45 * 10^{-9}} = 0.327 * 10^{-20}$ |
| $U_{9X_4} = \frac{A}{r_{9X}} = \frac{23.04 * 10^{-29}}{43.86 * 10^{-9}} = 0.525 * 10^{-20}$ | $U_{9Y_4} = \frac{A}{r_{9Y}} = \frac{23.04 * 10^{-29}}{58 * 10^{-9}} = 0.397 * 10^{-20}$ |
| $U_{10X_4} = \frac{A}{r_{10X}} = \frac{23.04 * 10^{-29}}{22 * 10^{-9}} = 1.047 * 10^{-20}$ | $U_{10Y_4} = \frac{A}{r_{10Y}} = \frac{23.04 * 10^{-29}}{43.86 * 10^{-9}} = 0.525 * 10^{-20}$ |
| $U_{11X_4} = \frac{A}{r_{11X}} = \frac{23.04 * 10^{-29}}{28.28 * 10^{-9}} = 0.814 * 10^{-20}$ | $U_{11Y_4} = \frac{A}{r_{11Y}} = \frac{23.04 * 10^{-29}}{53.74 * 10^{-9}} = 0.428 * 10^{-20}$ |
| $U_{12X_4} = \frac{A}{r_{12X}} = \frac{23.04 * 10^{-29}}{2.82 * 10^{-9}} = 8.17 * 10^{-20}$ | $U_{12Y_4} = \frac{A}{r_{12Y}} = \frac{23.04 * 10^{-29}}{28.28 * 10^{-9}} = 0.814 * 10^{-20}$ |
| $U_{13X_4} = \frac{A}{r_{13X}} = \frac{23.04 * 10^{-29}}{40 * 10^{-9}} = 0.576 * 10^{-20}$ | $U_{13Y_4} = \frac{A}{r_{13Y}} = \frac{23.04 * 10^{-29}}{60.72 * 10^{-9}} = 0.379 * 10^{-20}$ |
| $U_{14X_4} = \frac{A}{r_{14X}} = \frac{23.04 * 10^{-29}}{28.42 * 10^{-9}} = 0.810 * 10^{-20}$ | $U_{14Y_4} = \frac{A}{r_{14Y}} = \frac{23.04 * 10^{-29}}{40 * 10^{-9}} = 0.576 * 10^{-20}$ |
| $U_{15X_4} = \frac{A}{r_{15X}} = \frac{23.04 * 10^{-29}}{20 * 10^{-9}} = 1.152 * 10^{-20}$ | $U_{15Y_4} = \frac{A}{r_{15Y}} = \frac{23.04 * 10^{-29}}{42.04 * 10^{-9}} = 0.548 * 10^{-20}$ |
| $U_{16X_4} = \frac{A}{r_{16X}} = \frac{23.04 * 10^{-29}}{18.11 * 10^{-9}} = 1.272 * 10^{-20}$ | $U_{16Y_4} = \frac{A}{r_{16Y}} = \frac{23.04 * 10^{-29}}{20 * 10^{-9}} = 1.152 * 10^{-20}$ |
| $U_{17X_4} = \frac{A}{r_{17X}} = \frac{23.04 * 10^{-29}}{63.24 * 10^{-9}} = 0.364 * 10^{-20}$ | $U_{17Y_4} = \frac{A}{r_{17Y}} = \frac{23.04 * 10^{-29}}{78.02 * 10^{-9}} = 0.295 * 10^{-20}$ |
| $U_{18X_4} = \frac{A}{r_{18X}} = \frac{23.04 * 10^{-29}}{56.63 * 10^{-9}} = 0.406 * 10^{-20}$ | $U_{18Y_4} = \frac{A}{r_{18Y}} = \frac{23.04 * 10^{-29}}{63.24 * 10^{-9}} = 0.364 * 10^{-20}$ |
| $U_{19X_4} = \frac{A}{r_{19X}} = \frac{23.04 * 10^{-29}}{44.72 * 10^{-9}} = 0.519 * 10^{-20}$ | $U_{19Y_4} = \frac{A}{r_{19Y}} = \frac{23.04 * 10^{-29}}{58.03 * 10^{-9}} = 0.397 * 10^{-20}$ |
| $U_{20X_4} = \frac{A}{r_{20X}} = \frac{23.04 * 10^{-29}}{43.90 * 10^{-9}} = 0.524 * 10^{-20}$ | $U_{20Y_4} = \frac{A}{r_{20Y}} = \frac{23.04 * 10^{-29}}{44.72 * 10^{-9}} = 0.515 * 10^{-20}$ |